\documentclass[
prl, twocolumn%
,showpacs ,amssymb,superscriptaddress,aps
]{revtex4}
\usepackage{graphicx}
\input{epsf}

\usepackage{amsmath,amssymb}
\usepackage{bm}

\newcommand{\dalm}{\kern1pt\vbox{\hrule height 0.9pt\hbox{\vrule width 0.9pt
\hskip 2.5pt\vbox{\vskip 5.5pt}\hskip 3pt\vrule width 0.3pt}\hrule height 0.3pt}
\kern1pt}

\newcommand{\gsim}{\, \raisebox{-0.8ex}{$\stackrel{\textstyle >}{\sim}$ }}

\begin{document}

%\twocolumn[\hsize\textwidth\columnwidth\hsize\csname @twocolumnfals\endcsname

% For two column
%\wideabs{

\title{Probing the Equation of State of Nuclear Matter via 
Neutron Star Asteroseismology}

%\title{Probing crust EOS via asteroseismology}

\author{Hajime Sotani}
\email{hajime.sotani@nao.ac.jp}
\affiliation{National Astronomical Observatory of Japan, 2-21-1 Osawa, Mitaka, Tokyo 
181-8588, Japan
}

\author{Ken'ichiro Nakazato}
\affiliation{Faculty of Science \& Technology, Tokyo University of Science, 2641 Yamazaki, Noda, Chiba 278-8510, Japan
}

\author{Kei Iida}
\affiliation{Department of Natural Science, Kochi University, 2-5-1 Akebono-cho, Kochi 780-8520, Japan
}

\author{Kazuhiro Oyamatsu}
\affiliation{Department of Human Informatics, Aichi Shukutoku University,
9 Katahira, Nagakute, Aichi 480-1197, Japan
%$^1$National Astronomical Observatory of Japan, Osawa, Mitaka, Tokyo 
%181-8588, Japan \\
%$^2$Department of Physics, Tokyo University of Science, Yamazaki, Noda, Chiba 
%278-8510, Japan \\
%$^2$Faculty of Science \& Technology, Tokyo University of Science, Yamazaki 2641, Noda, Chiba 278-8510, Japan \\
%$^3$Department of Natural Science, Kochi University, Akebono-cho, Kochi 
%780-8520, Japan \\
%$^4$Department of Human Informatics, Aichi Shukutoku University,
%Nagakute, Nagakute-cho, Aichi-gun, Aichi, 480-1197, Japan
%$^1$Division of Theoretical Astronomy, National Astronomical Observatory of Japan, 2-21-1 Osawa, Mitaka, Tokyo 181-8588, Japan \\
%$^2$Department of Physics, Faculty of Science \& Technology, Tokyo University of Science, Yamazaki 2641, Noda, Chiba 278-8510, Japan \\
%$^3$Department of Natural Science, Kochi University, 2-5-1 Akebono-cho, Kochi 780-8520, Japan \\
%$^4$Department of Human Informatics, Aichi Shukutoku University 9 Katahira Nagakute, Nagakute-cho, Aichi-gun, Aichi, 480-1197, Japan
}

\date{\today}

% Abstract
\begin{abstract}
We general relativistically calculate the frequency of fundamental torsional 
oscillations of neutron star crusts, where we focus on the crystalline properties
obtained from macroscopic nuclear models in a way depending on the equation of 
state of nuclear matter. We find that the calculated frequency is sensitive 
to the density dependence of the symmetry energy, but almost independent of 
the incompressibility of symmetric nuclear matter.  By identifying the 
lowest-frequency quasi-periodic oscillation in giant flares observed from soft 
gamma-ray repeaters as the fundamental torsional mode and allowing for the 
dependence of the calculated frequency on stellar models, we provide a lower 
limit of the density derivative of the symmetry energy as $L\simeq 50$ MeV.

%The equation of state of nuclear matter is becoming to reveal via the experiments, but there is still some uncertainty in the parameters. In order to make a constraint to these parameters, we examine the torsional oscillations of compact objects and compare with the observed evidences in the giant flares, which is completely different technique from the ground experiments to restrict the equation of state. We find that the frequencies of fundamental torsional oscillations are almost independent from the incompressibility of symmetric nuclear matter if the stellar mass and radius are fixed. Compared with the observations in the giant flares, one can expect that the symmetry energy density derivative coefficient should be more than 50 MeV independent of the stellar models.
\end{abstract}

\pacs{04.40.Dg, 21.65.Ef, 26.60.Gj, 97.10.Sj}
%
%%%%%%%%%%%%%%%%%%%%%%%%%%%%%%%%%%%%%%%%%%%%%%%%%
%  04.40.Dg :  Relativistic stars: structure, stability, and oscillations (see also 97.60.-s Late stages of stellar evolution) 
%  21.65.Ef  :  Symmetry energy
%  26.60.Gj  :  Neutron star crust
%  21.60.-n  :   Nuclear structure models and methods
%  97.10.Sj  :   Pulsations, oscillations, and stellar seismology 
%%%%%%%%%%%%%%%%%%%%%%%%%%%%%%%%%%%%%%%%%%%%%%%%%
%]
% For two column
%}
\maketitle
%\baselineskip 24pt
%%%%%%%%%%%%%%%%%%%%%%%%%%%%%%%%%%%%%%%%%%%%%%%%
%\section{Introduction}
%\label{sec:I}
%%%%%%%%%%%%%%%%%%%%%%%%%%%%%%%%%%%%%%%%%%%%%%%%

    Observations of the global oscillations of stars play a significant role 
in probing the properties of stellar matter, as in the case of the Sun.  
Studies in this direction are often referred to as asteroseismology. 
Oscillations of neutron stars are expected to give an insight to the properties
of matter under conditions of ultra-high density.  Recent observations of the 
quasi-periodic oscillations (QPOs) in giant flares from soft gamma-ray 
repeaters (SGRs) \cite{WS2006} suggest that a long-awaited evidence for the 
neutron star oscillations has been given with the help of coupling of gigantic 
magnetic fields with the solid crust.  So far, three giant flares have 
been detected from SGR 0526-66, SGR 1900+14, and SGR 1806-20, and, through the 
timing analysis of the X-ray afterglow, the QPOs with frequencies in the range 
from tens Hz up to a few kHz have been discovered \cite{WS2006}.  Many 
theoretical attempts to explain the observed QPO frequencies have been done in 
terms of the torsional oscillations in the crustal region and/or the magnetic 
oscillations (e.g., Refs.\  
\cite{Levin2006,Lee2007,SA2007,Sotani2007,Sotani2008b,Sotani2009,
GCFMS2011,CK2011}).  If the QPOs in giant flares are associated with
the crustal oscillations, the properties of inhomogeneous nuclear matter in 
the crust could be clarified \cite{SW2009,Sotani2011b,GNHL2011}.

%One of the most promissing evidences about the oscillations of neutron stars is the quasi-periodic oscillations (QPOs) in giant flares. The soft gamma repeaters (SGRs) are considered as the candidate of magnetars, which are neutron stars with strong magnetic fields \cite{DT1992}. In addition to the sporadic X- and gamma-ray bursts, SGRs rarely emit much stronger gamma-rays called ``giant flares". Up to now, at least three giant flares have been detected, i.e., SGR 0526-66, SGR 1900+14, and SGR 1806-20. Through the timing analysis of the decaying tail of giant flares, the existence of QPOs has discovered, whose frequencies are in the range from tens Hz up to a few kHz \cite{WS2006}. To explain these observed QPO frequencies theoretically, many attempts have done with the torsional oscillations in the crust region and/or the magnetic oscillations (e.g., \cite{Levin2006,Lee2007,SA2007,Sotani2007,Sotani2008a,Sotani2008b,Sotani2009,GCFMS2011,CK2011}). Associated with the QPOs in giant flares, there also exist the suggestions of the importance to consider the microscopic properties of crust region \cite{SW2009,W2011} and the possibility to obtain the information about nonuniform nuclear structure in the crust \cite{Sotani2011b,GNHL2011}.

    The outer part of neutron stars can be described as follows. Below an
ionic ocean in the vicinity of the surface, a bcc Coulomb lattice of nuclei 
embedded in a roughly uniform electron gas is considered to compose a 
crustal region.  In this region, nuclei become gradually neutron-rich with 
increasing density and even drip neutrons at a density of about
$4\times10^{11}$ g cm$^{-3}$.  Near normal nuclear density, this crustal 
region is considered to melt into uniform nuclear matter.  Just before 
melting, roughly spherical nuclei are so closely packed that global 
deformations into rodlike nuclei could occur.  As the density increases
further, possible changes of the nuclear shape are rod, slab, tube, and bubble 
\cite{LRP1993,O1993}.  These exotic nuclei are often called ``nuclear pasta.''
Not only the density region of the pasta phases but also the charge number of 
roughly spherical nuclei is known to be sensitive to the empirically uncertain 
density dependence of the symmetry energy \cite{OI2007,NGL2011}.
Generally, it is difficult to observationally probe the properties of matter in
the crust, but the asteroseismology could exceptionally help to constrain the 
matter properties such as nuclei present and the equation of state (EOS) (e.g.,
Refs.\ \cite{AK1996,Sotani2001,Sotani2004,SYMT2011}).  In this Letter, we will 
show by systematic analyses that an approach to the QPOs in SGR giant flares 
in terms of the torsional shear modes in the crust could severely constrain 
the density dependence of the symmetry energy.

    We begin with the bulk energy per nucleon near the saturation point of 
symmetric nuclear matter at zero temperature, which can be written as a 
function of nucleon density $n$ and neutron excess $\alpha$ \cite{L1981}:
\begin{equation}
w = w_0  + \frac{K_0}{18n_0^2}(n-n_0)^2 + \left[S_0 
         + \frac{L}{3n_0}(n-n_0)\right]\alpha^2, \label{eq:w}
\end{equation}
where $w_0$, $n_0$, and $K_0$ are the saturation energy, saturation density, 
and incompressibility of symmetric nuclear matter, respectively.  The 
parameters $L$ and $S_0$ characterize the symmetry energy coefficient $S(n)$:
$S_0=S(n_0)$ is the symmetry energy coefficient at $n=n_0$, while 
$L=3n_0(dS/dn)_{n=n_0}$ is the symmetry energy density derivative coefficient.
The parameters $w_0$, $n_0$, and $S_0$ can be relatively easier to determine 
from empirical masses and radii of stable nuclei \cite{OI2003}.  On the other 
hand, the remaining two parameters, $L$ and $K_0$, are more difficult to 
determine.  Here we introduce a new parameter, $y$, defined as 
$y=-K_0S_0/(3n_0L)$, which denotes the slope of the saturation line in the 
vicinity of $\alpha=0$ \cite{OI2003}.  Two of us (K.O. and K.I.) constructed 
the model for the EOS of nuclear matter in such a way as to reproduce Eq.\ 
(\ref{eq:w}) in the limit of $n\to n_0$ and $\alpha\to0$, calculated the 
optimal density distribution of stable nuclei within a simplified version of 
the extended Thomas-Fermi theory, and obtained the values of $w_0$, $n_0$, and 
$S_0$ for given $y$ (or $L$) and $K_0$ by fitting the charge number, mass 
excess, and charge radius that can be calculated from the optimal density 
distribution to the empirical behavior.  In a manner similar to Ref.\ 
\cite{OI2007}, we adopt the parameter range satisfying $0< L< 160$ MeV, 180 MeV
$\le K_0 \le$ 360 MeV, and $y< -200$ MeV fm$^3$.  In fact, this parameter range
equally well reproduces the mass and radius data for stable nuclei and 
effectively covers even extreme cases \cite{OI2003}.  With such parameters, we 
can obtain the crust EOS and the equilibrium nuclear shape and size by 
generalizing the Thomas-Fermi model for nuclei to matter in the crust.  The 
results for the nucleon densities $n_1$ at which the nuclear shape changes from
spherical to cylindrical one and $n_2$ at which the nuclear matter becomes 
uniform were tabulated for several sets of $y$ (or $L$) and $K_0$ in Table 
\ref{tab:EOS}.  Note that the interval between $n_1$ and $n_2$, which 
corresponds to the pasta region, decreases with $L$ and vanishes at $L\sim100$
MeV.

\begin{table}[t]
\begin{center}
\leavevmode
\caption{The pasta density region calculated for several sets of the EOS 
parameters.
%Crust EOS parameters
}
\begin{tabular}{ccccccc}
\hline\hline
 & $y$ (MeV fm$^3$) & $K_0$ (MeV) & $L$ (MeV) & $n_1$ (fm$^{-3}$) & $n_2$ (fm$^{-3}$) & \\
\hline
 &   $-220$ & 180 & 52.2 & 0.060 & 0.079 &  \\
 &   $-220$ & 230 & 73.4 & 0.064 & 0.073 &  \\
%  &   -220 & 280 & *** & *** & *** &  \\
 &   $-220$ & 360 & 146.1 & 0.066 & 0.066 &  \\
 &   $-350$ & 180 & 31.0 & 0.058 & 0.091 &  \\
 &   $-350$ & 230 & 42.6 & 0.063 & 0.086 &  \\
%  &   -350 & 280 & *** & *** & *** &  \\
 &  $ -350$ & 360 & 76.4 & 0.072 & 0.076 &  \\
 & $-1800$ & 180 & 5.7 & 0.058 & 0.134 &  \\
 & $-1800$ & 230 & 7.6 & 0.058 & 0.127 &  \\
%  & -1800 & 280 & *** & *** & *** &  \\
 & $-1800$ & 360 & 12.8 & 0.058 & 0.118 &  \\
\hline\hline
\end{tabular}
\label{tab:EOS}
\end{center}
\end{table}
%%%%%%%%%%%%%%%%%%%%%%%%%%%%%%%%%%%

    We turn to the equilibrium structure of nonrotating neutron stars, which
is determined by the Tolman-Oppenheimer-Volkoff (TOV) equations for given EOS 
of neutron star matter.  For a solution to these equations, the metric can be 
described in terms of the spherical polar coordinates $r$, $\theta$, and $\phi$
as
\begin{equation}
 ds^2 = -{\rm e}^{2\Phi(r)}dt^2 + {\rm e}^{2\Lambda(r)}dr^2 + r^2 d\theta^2 + r^2\sin^2\theta\, d\phi^2.
\end{equation}
%where $\Phi$ and $\Lambda$ are functions of $r$. 
However, the EOS for the core
surrounded by the crust is still uncertain, although there are many 
calculations based on realistic nuclear interactions and even constraints from 
the observationally deduced neutron star masses and radii 
\cite{OBG2010,SLB2010}.  To avoid this uncertainty, we here construct the crust
with the EOS models mentioned above by solving the TOV equations inward from 
the surface of the star for given stellar mass $M$ and radius $R$ as in Ref.\ 
\cite{IS1997}.  We remark that the thickness of the crust thus constructed is 
consistent with a typical behavior given in terms of $M$ and $R$ by Eq.\ 
(18) in Ref. \cite{RP1994}.

   We next consider the shear modulus of a crustal part composed of spherical 
nuclei of charge $Ze$ and number density $n_i$, which can be approximately 
described as \cite{SHOII1991}
\begin{equation}
\mu = 0.1194 n_i (Ze)^2/a, \label{eq:shear}
\end{equation}
where $a = (3/4\pi n_i)^{1/3}$ is the Wigner-Seitz radius.  Note that this 
formula is derived in the limit of zero temperature from Monte Carlo 
calculations of the shear modulus averaged over all directions for a perfect 
bcc Coulomb crystal of point charges embedded in a neutralizing uniform 
background \cite{OI1990}.  We will use this formula for calculations of the torsional 
oscillation frequencies since effects of quantum zero-point motions and thermal
fluctuations are negligible.  For pasta nuclei except bubbles, the elasticity 
is expected to be much lower than that for spherical nuclei \cite{PP1998}.  
Moreover, the core, which is expected to be composed mostly of fluids, may 
have a structure with nonvanishing elasticity \cite{JO2011}.  Here, we simply 
assume $\mu=0$ at $n>n_1$, as in Ref.\ \cite{GNHL2011}.  Within this 
assumption, the shear modulus at $n>n_1$ is underestimated, which tends to 
lower the torsional oscillation frequencies.  The frequencies as will be 
estimated below should thus be regarded as lower limits basically, but still 
will be shown to play a role in constraining $L$.

%In the limit of zero temperature, the shear modulus can be described as
%\begin{equation}
% \mu = 0.1194 \times n_i (Ze)^2/a, \label{eq:shear}
%\end{equation}
%where $n_i$ and $+Ze$ are corresponding to the ion number density and the ion charge, while $a$ is defined as $a^3 = 3/(4\pi n_i)$, which corresponds the average ion spacing \citep{SHOII1991}. It should be noticed that this approximate formula for the shear modulus is derived on the assumption of bcc crystalline state, which is averaged over all directions \cite{note}. So, if one takes into account the effect of nonuniform nuclear structure known as nuclear pasta, the shear modulus might be modified \cite{Sotani2011b}. Actually, it is suggested that the elastic properties in pasta phase could be softer than that in bcc lattice \cite{PP1998}. However, since the realistic shear modulus in pasta phase is not revealed yet, in this article we will deal with the pasta phase as liquid, i.e., $\mu=0$ for $n\ge n_1$, which is the same treatment as in \cite{GNHL2011}. With this assumption, the shear modulus is estimated smaller effectively. As a result, the corresponding frequencies of torsional oscillations also become smaller, i.e., the obtained frequencies in this article could be corresponding to lower limit. But, as mentioned latter, the effect of pasta phase might not be so important in the torsional oscillations.

    Since the torsional oscillations on a spherically symmetric star are 
incompressible, the star is free from deformation and density variation during
such oscillations.  One can thus determine the frequencies of the torsional 
oscillations with satisfactory accuracy even if one neglects the resulting
metric perturbations by setting $\delta g_{\mu\nu}=0$, which is known as the 
relativistic Cowling approximation.  Within this approximation, the torsional 
oscillations can be described by a single perturbation variable, i.e., the 
angular displacement of the stellar matter, ${\cal Y}$, which is related to
the $\phi$-component of the perturbed 4-velocity of a matter element, 
$\delta u^{\phi}$, by $\delta u^\phi = {\rm e}^{-\Phi}\partial_t {\cal Y}(t,r) 
\partial_\theta P_\ell (\cos\theta)/\sin{\theta}$ with the $\ell$-th order  
Legendre polynomial $P_\ell$.  Assuming
${\cal Y}(t,r)={\rm e}^{{\rm i}\omega t}{\cal Y}(r)$, one can obtain the 
perturbation equation for ${\cal Y}(r)$ from the linearized equation of 
motion as \cite{ST1983}
\begin{align}
 {\cal Y}'' &+ \left[\left(\frac{4}{r}+\Phi'-\Lambda'\right)+\frac{\mu'}{\mu}\right]{\cal Y}'  \nonumber \\
   &+ \left[\frac{\epsilon+p}{\mu}\omega^2{\rm e}^{-2\Phi}-\frac{(\ell+2)(\ell-1)}{r^2}\right]{\rm e}^{2\Lambda}{\cal Y} = 0,
 \label{eq:perturbation}
\end{align}
where $\epsilon$ and $p$ are the energy density and pressure, and the prime 
denotes the derivative with respect to $r$.  To determine the eigenfrequencies,
we adopt the zero-traction condition at $n=n_1$ and the zero-torque condition 
at the star's surface.  These boundary conditions reduce to ${\cal Y}'=0$ at 
$n=0, n_1$ \cite{ST1983,Sotani2007}.  We remark that neutron superfluidity 
ignored here would subtract the mass density of superfluid neutrons from the 
enthalpy density $\epsilon+p$ and hence enhance the eigenfrequencies 
\cite{PCR2010}.

\begin{figure}[t]
\begin{center}
\includegraphics[scale=0.425]{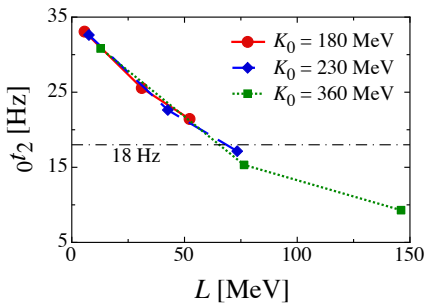} 
\end{center}
\caption{%%
(Color online) Frequency of the $\ell=2$ fundamental torsional oscillations, 
${}_0t_2$, plotted as a function of $L$ for $M=1.4M_\odot$ and $R=12$ km. 
%The circles, diamonds, and squares are the results for $K_0=180$, 230, and 360 MeV.
The horizontal dot-dashed line denotes the lowest QPO 
frequency observed from SGR 1806-20 \cite{WS2006}.
%Frequencies of $\ell=2$ fundamental torsional oscillations, ${}_0t_2$, are plotted as a function of $L$, where $M=1.4M_\odot$ and $R=12$km. The horizontal dot-dashed line denotes the lowest QPO frequency in the SGR 1806-20.
}%%
\label{fig:0t2M14R12}
\end{figure}

    First, to see the dependence of the fundamental torsional oscillations on 
the EOS parameters, we calculate the corresponding eigenfrequency for a 
typical stellar model with $M=1.4M_\odot$ and $R=12$ km.  Figure 
\ref{fig:0t2M14R12} shows the frequency of the $\ell=2$ fundamental torsional 
oscillations, $_0t_2$, calculated for the nine sets of $L$ and $K_0$ that are
tabulated in Table \ref{tab:EOS}, together with the lowest QPO frequency in SGR
1806-20 \cite{WS2006}.  From this figure, one can observe that $_0t_2$ is 
almost independent of $K_0$ in the parameter range adopted here once the 
stellar model is fixed.  Such independence can be seen for various stellar 
models ranging $R=10$, $12$, and $14$ km as well as $M=1.4M_\odot$ and 
$1.8M_\odot$.  We can thus focus on the $L$ dependence of the calculated 
$_0t_2$, which arises mainly because the nuclear charge $Z$ decreases with $L$ 
through the surface property \cite{OI2007}, leading to decrease in the shear 
modulus (\ref{eq:shear}) with $L$.
Since $Z$ depends strongly on $L$ in the vicinity of $n=n_1$, this region is
likely to play a role in constraining $L$ via the evaluations of $_0t_2$.
To see the $L$ dependence explicitly, we 
derive a fitting formula for $_0t_2$ as
\begin{equation}
 {}_0t_2 = c_0 - c_1 \frac{L}{100\ {\rm MeV}} + c_2 \left(\frac{L}{100\ {\rm MeV}}\right)^2, \label{eq:emp}
\end{equation}
where $c_0$, $c_1$, and $c_2$ are the adjustable parameters that depend on 
$M$ and $R$.  The values of these parameters for $M=1.4M_\odot$ are shown in 
Table \ref{tab:Empirical}.

%First, to see the dependence of the fundamental torsional oscillations on the EOS parameters, we calculate such frequencies on the typical stellar model with $M=1.4M_\odot$ and $R=12$km. Fig. \ref{fig:0t2M14R12} shows the frequencies of $\ell=2$ fundamental torsional oscillations as a function of $L$, where the lines with circles, diamonds, and squares are corresponding to the results for $K_0=180$, 230, and 360 MeV. In addition to the numerical results, for comparison, the dot-dashed line denotes the lowest observed frequency in the SGR 1806-20, which is 18 Hz. From this figure, one can observe that the frequencies of fundamental torsional oscillations are almost independent from the value of $K_0$ at least in the parameter range of our calculations, if the stellar model is fixed. We make sure of this dependence with different stellar models, such as $R=10$, $12$, and $14$km for $M=1.4M_\odot$ and $1.8M_\odot$. It could be because the charge number of spherical nuclei, Z, which contributes largely to the shear modulus as in Eq. (\ref{eq:shear}), depends strongly on not $K_0$ but $L$ \cite{OI2007}. As a consequence, one can derive the fitting formula of the frequencies of $\ell=2$ fundamental torsional oscillations as
%\begin{equation}
%  {}_0t_2 = c_0 - c_1 L + c_2 L^2, \label{eq:emp}
%\end{equation}
%where $c_0$, $c_1$, and $c_2$ are constants depending on the stellar properties. The examples of these parameters are shown in Table \ref{tab:Empirical}. 

%%%%%%%%%%%%%%%%%%%%%%%%%%%%%%%%%%%
% Figure 1
%%%%%%%%%%%%%%%%%%%%%%%%%%%%%%%%%%%
%%%%%%%%%%%%%%%%%%%%%%%%%%%%%%%%%%%

%%%%%%%%%%%%%%%%%%%%%%%%%%%%%%%%%%%
% Table 2
%%%%%%%%%%%%%%%%%%%%%%%%%%%%%%%%%%%
\begin{table}[t]
\begin{center}
\leavevmode
\caption{$\chi^2$ fitting via formula (\ref{eq:emp}) for $M=1.4M_\odot$,
where $\chi^2$ is given for the deviation of the calculated ${_0t_2}$ from the fitting formula (\ref{eq:emp}).
%Coefficients in fitting formula (\ref{eq:emp}) for $M=1.4M_\odot$.
}
\begin{tabular}{ccccccc}
\hline\hline
 & $R$ (km) & $c_0$ (Hz) & $c_1$ (Hz) & $c_2$  (Hz) & $\chi^2$ & \\
\hline
 & 10 & 38.95 & 36.14 & 11.34 & 1.944  &  \\
 & 12 & 34.87 & 32.33 & 10.14 & 1.539  &  \\
 & 14 & 31.48 & 29.14 & 9.123 & 1.239  &  \\
\hline\hline
\end{tabular}
\label{tab:Empirical}
\end{center}
\end{table}
%%%%%%%%%%%%%%%%%%%%%%%%%%%%%%%%%%%

\begin{figure}[t]
\begin{center}
\includegraphics[scale=0.425]{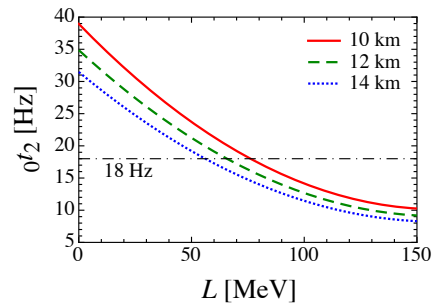} 
\end{center}
\caption{%%
(Color online) ${}_0t_2$ given by formula (\ref{eq:emp}) for $M=1.4M_\odot$ 
and $R=10, 12, 14$ km.  
%Frequencies of $\ell=2$ fundamental torsional oscillations expected from the empirical formula are plotted as a function of $L$, where $M=1.4M_\odot$
}%%
\label{fig:0t2M14}
\end{figure}

\begin{figure}[t]
\begin{center}
\includegraphics[scale=0.425]{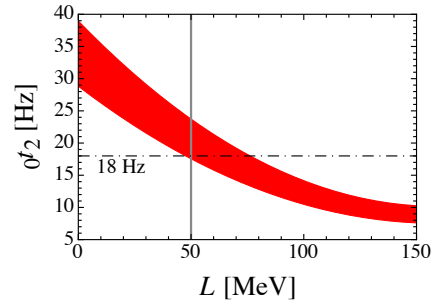} 
\end{center}
\caption{%%
(Color online) Same as Fig.\ \ref{fig:0t2M14} for $10$ km $\le R \le14$ km and 
$1.4M_\odot\le M\le 1.8M_\odot$.
}%%
\label{fig:0t2}
\end{figure}

    The fitting formula thus obtained is exhibited in Fig.\ \ref{fig:0t2M14}. 
We find from this figure that $_0t_2$ for fixed $L$ can be determined within the 
accuracy of $\sim20\%$, if $R$ is in the range of $10$-$14$ km.  By shifting 
$M$ from $1.4M_\odot$ up to $1.8M_\odot$, we find that $_0t_2$ for $R=10$, 12, 
and 14 km decreases only by $\sim 14$, 10, and 9$\%$, respectively.  For 
clarity, we plot, in Fig.\ \ref{fig:0t2}, $_0t_2$ estimated for stellar models 
ranging $10$ km $\le R \le 14$ km and $1.4M_\odot \le M \le 1.8M_\odot$.  The 
results are confined within the painted region, which is so narrow that we can
constrain $L$ as we shall see.

%With the obtained fitting formula, one can make a figure as Fig. \ref{fig:0t2M14}, where the stellar mass is fixed to be $M=1.4M_\odot$. It is found from this figure that the dependence of the frequency of $\ell=2$ fundamental torsional oscillation on the stellar radius is relatively small. Namely, the frequency with the fixed value of $L$ is determined within the accuracy of $\sim20\%$, if the stellar radius is in the range of $10\le R\le 14$km. Additionally, we find that the dependence of ${}_0t_2$ on the stellar mass becomes smaller than that on the radius, i.e., the frequencies for $R=10$, 12, and 14 km decrease $\sim 14$, 10, and 9$\%$ if the stellar mass increases from $1.4M_\odot$ up to $1.8M_\odot$. In practice, the frequencies of $\ell=2$ fundamental torsional oscillations for the stellar models with $10$ km $\le R \le 14$ km and $1.4M_\odot \le M \le 1.8M_\odot$ can be confined in the colored region in Fig. \ref{fig:0t2}. Thus, almost independent of the stellar models, one could be possible to restrict on the value of $L$ via the observations of torsional oscillations.

%%%%%%%%%%%%%%%%%%%%%%%%%%%%%%%%%%%
% Figure 2
%%%%%%%%%%%%%%%%%%%%%%%%%%%%%%%%%%%
%%%%%%%%%%%%%%%%%%%%%%%%%%%%%%%%%%%

%%%%%%%%%%%%%%%%%%%%%%%%%%%%%%%%%%%
% Figure 3
%%%%%%%%%%%%%%%%%%%%%%%%%%%%%%%%%%%
%%%%%%%%%%%%%%%%%%%%%%%%%%%%%%%%%%%

    Let us assume that the observed QPOs in SGR giant flares arise from the 
torsional oscillations in neutron star crusts and note that among many 
eigenfrequencies of the torsional oscillations, $_0t_2$ is the lowest.  Then, 
$_0t_2$ would become equal to or even lower than the lowest frequency in the 
observed QPOs.  Consequently, one can constrain $L$ as $L\gsim 50$ MeV from 
the painted region in Fig.\ \ref{fig:0t2}.  Recall that the calculated 
$_0t_2$ is likely to be underestimated because of the simplified treatments 
of the shear modulus and the enthalpy density.  In particular, elasticity in 
the pasta phases ignored here would act to increase $_0t_2$ as long as 
$n_2>n_1$ \cite{GNHL2011}.  At $L\gsim50$ MeV, however, $n_2-n_1$ is 
already small (see Table \ref{tab:EOS}).  The resultant modifications on 
the $L$ constraint are thus expected to be small.

%It is considered that the observed QPOs in giant flares can be due to the torsional oscillations in neutron star crust. On the other hand, among many frequencies of torsional oscillations, the frequency of $\ell=2$ fundamental oscillations is the lowest one. So, if the observed QPO frequencies would be corresponding to the torsional oscillations in neutron star crust, the frequency of $\ell=2$ fundamental torsional oscillation should become lower than the lowest frequency in the observed QPOs. Based on this statement, one can restrict on $L$ as $L\gsim 50$ MeV independent of the stellar models (see Fig. \ref{fig:0t2}). It should be noticed that as mentioned the above, the calculated frequencies might be underestimated because we neglect the effect of pasta phase in our calculations. In other words, the restriction that $L\gsim 50$ MeV might be severer than that including the effect of pasta phase. However, the region of past phase become narrower as $L$ increases, and that region could vanish for $L\gsim 100$ MeV \cite{OI2007}. Thus, even if the effect of pasta phase will be taken into account, the restriction that $L\gsim 50$ MeV may not be modified.

%%%%%%%%%%%%%%%%%%%%%%%%%%%%%%%%%%%
% Figure 4
%%%%%%%%%%%%%%%%%%%%%%%%%%%%%%%%%%%
\begin{figure}[t]
\begin{center}
\includegraphics[scale=0.425]{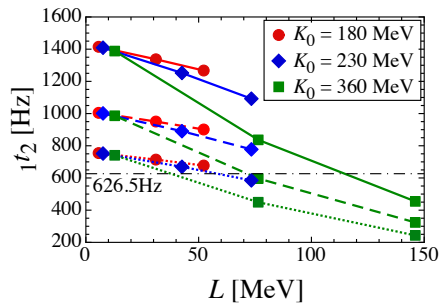} 
\end{center}
\caption{%%
(Color online) Frequency of the first overtone of the $\ell=2$ torsional 
oscillations, $_1t_2$, plotted as a function of $L$ for $M=1.4M_\odot$.  The 
solid, broken, and dotted lines are the results for $R=10$, 12, and 14 km,
while the circles, diamonds, and squares are those for $K_0=180$, 230, and 
360 MeV.  The horizontal dot-dashed line denotes one of the high QPO 
frequencies observed from SGR 1806-20 \cite{WS2006}.
%(Color online) The 1st overtone frequencies of $\ell=2$ torsional oscillations, ${}_1t_2$, are plotted as a function of $L$, where $M=1.4M_\odot$. The solid, broken, and dotted lines correspond to the frequencies for the stellar models with $R=10$, 12, and 14 km. The horizontal dot-dashed line denotes the higher QPO frequency in the SGR 1806-20.
}%%
\label{fig:0t2M18}
\end{figure}
%%%%%%%%%%%%%%%%%%%%%%%%%%%%%%%%%%%

    We proceed to show the frequency of the first overtone of the $\ell=2$ 
torsional oscillations, $_1t_2$, calculated again for the nine sets of $L$ and 
$K_0$ that are tabulated in Table \ref{tab:EOS}.  The results with 
$M=1.4M_\odot$ and $R=10$, 12, 14 km are plotted in Fig.\ \ref{fig:0t2M18}.
We find from this figure that $_1t_2$ depends not only on $L$, but on $K_0$ 
and $R$ significantly, in contrast to the case of $_0t_2$.  The $R$ dependence 
arises because basically, $_1t_2$ is inversely proportional to the crust 
thickness, which in turn increases as $R^2$ \cite{RP1994}.  The $K_0$ 
dependence comes presumably from the $K_0$ dependence of $n_1$ (see Table 
\ref{tab:EOS}), but could be modified drastically once elasticity in the pasta
phases is allowed for.  This issue will be addressed elsewhere \cite{SNIO}.

%Furthermore, we also find that, unlike the fundamental torsional oscillations, the 1st overtone frequencies depend strongly on $K_0$ and $L$ (see Fig. \ref{fig:0t2M18}). This could be because the overtone frequencies are associated with the crust thickness, which depends on  $K_0$. Therefore, with the restriction on $L$ via the fundamental frequency, one could be also possible to make a constraint in $K_0$ via the observations of overtone.

%%%%%%%%%%%%%%%%%%%%%%%%%%%%%%%%%%%%%%%%%%%%%%%%
%\section{Conclusion}
%\label{sec:V}
%%%%%%%%%%%%%%%%%%%%%%%%%%%%%%%%%%%%%%%%%%%%%%%%

   In summary, we have investigated the fundamental torsional 
mode and the first overtone in neutron star crusts for various EOS and stellar
models.  The identification of the lowest QPO frequency observed from SGR 
1806-20 as $_0t_2$ would then allow us to constrain $L$ as $L\gsim50$ MeV.  
At present, this constraint is fairly stringent because experimental 
constraints on $L$ have yet to converge \cite{exp}.  While there are earlier 
publications that remark the sensitivity of $_0t_2$ and $_1t_2$ to $L$ 
\cite{SW2009,GNHL2011}, the present work is the first to provide a 
lower limit of $L$ by sufficiently accurate and systematic calculations through the
general relativistic mode evaluations and the Thomas-Fermi treatment of nuclei. 
The present constraint already suggests that the pasta phases, if any in neutron 
stars, would occur in a narrow density region.  Neutron superfluidity and 
pasta elasticity would make the constraint on $L$ and the pasta region even 
more stringent.

\acknowledgments
%%%%%%%%%%%%%%%%%%%%%%%%%%%%%%%%%%%%%%%%%%%%%%%%
    H.S. is grateful to T. Tatsumi for fruitful discussions. This work 
was supported in part by Grants-in-Aid for Scientific Research on Innovative 
Areas through No.\ 23105711 and on Research Activity Start-up through No.\
23840038, which were provided by the Ministry of Education, Culture, Sports, 
Science, and Technology of Japan.

%HS is grateful to T. Tatsumi for his fruitful discussions. This work was supported by Grant-in-Aid for Scientific Research on Innovative Areas (23105711).

%\appendix
%%%%%%%%%%%%%%%%%%%%%%%%%%%%%%%%%%%%%%%%%%%%%%%%
%\section{}   % Appendix A
%\label{sec:appendix_1}
%%%%%%%%%%%%%%%%%%%%%%%%%%%%%%%%%%%%%%%%%%%%%%%%

%%%%%%%%%%%%%%%%%%%%%%%%%%%%%%%%%%%%%%%%%%%%%%%%


\begin{thebibliography}{999}
%%%%%%%%%%%%%%%%%%%%%%%%%%%%%%%%%%%%%%%%%%%%%%%%

%\bibitem{DT1992}
%    R. C. Duncan and C. Thompson, Astrophys.\ J. {\bf 392}, L9 (1992).

\bibitem{WS2006}
   A. L. Watts and T. E. Strohmayer, Adv.\ Space Res.\ {\bf 40}, 1446 (2006).

\bibitem{Levin2006}
   Y. Levin, Mon.\ Not.\ R. Astron.\ Soc.\ {\bf 368}, L35 (2006).

\bibitem{Lee2007}
   U. Lee, Mon.\ Not.\ R. Astron.\ Soc.\ {\bf 374}, 1015 (2007).

\bibitem{SA2007}
   L. Samuelsson and N. Andersson, Mon.\ Not.\ R. Astron.\ Soc.\ {\bf 374}, 256 (2007).

\bibitem{Sotani2007}
   H. Sotani, K. D. Kokkotas, and N. Stergioulas, Mon.\ Not.\ R. Astron.\ Soc.\ {\bf 375}, 261(2007); {\bf 385}, L5 (2008).

%\bibitem{Sotani2008a}
%   H. Sotani, K. D. Kokkotas, and N. Stergioulas, Mon.\ Not.\ R. Astron.\ Soc.\ {\bf 385}, L5 (2008).

\bibitem{Sotani2008b}
   H. Sotani, A. Colaiuda, and K. D. Kokkotas, Mon.\ Not.\ R. Astron.\ Soc.\ {\bf 385}, 2161 (2008).

\bibitem{Sotani2009}
   H. Sotani and K. D. Kokkotas, Mon.\ Not.\ R. Astron.\ Soc.\ {\bf 395}, 1163 (2009).

\bibitem{GCFMS2011}
   M. Gabler {\it et al.}, Mon.\ Not.\ R. Astron.\ Soc.\ {\bf 410}, L37 (2011).

\bibitem{CK2011}
   A. Colaiuda and K. D. Kokkotas,  Mon.\ Not.\ R. Astron.\ Soc.\ {\bf 414}, 3014 (2011).

%\bibitem{CK20112}
%   A. Colaiuda and K. D. Kokkotas,  arXiv:1112.3561.

\bibitem{SW2009}
   A. W. Steiner and A. L. Watts, Phys.\ Rev.\ Lett.\ {\bf 103}, 181101 (2009).

%\bibitem{W2011}
%    A. L. Watts, arXiv:1111.0514.

\bibitem{Sotani2011b}
   H. Sotani, Mon.\ Not.\ R. Astron.\ Soc.\ {\bf 417}, L70 (2011).

\bibitem{GNHL2011}
   M. Gearheart, W. G. Newton, J. Hooker, and B. A. Li, arXiv:1106.4875.

\bibitem{LRP1993}
   C. P. Lorenz, D. G. Ravenhall, and C. J. Pethick, Phys.\ Rev.\ Lett.\ {\bf 70}, 379 (1993).

\bibitem{O1993}
   K. Oyamatsu, Nucl.\ Phys.\ A {\bf 561}, 431 (1993).

%\bibitem{SOT1995}
%   K. Sumiyoshi, K. Oyamatsu, and H. Toki, Nucl.\ Phys.\ A {\bf 595}, 327 (1995).

%\bibitem{WM2011}
%    G. Watanabe and T. Maruyama, arXiv:1109.3511

\bibitem{OI2007}
   K. Oyamatsu and K. Iida, Phys.\ Rev.\ C {\bf 75}, 015801 (2007).

\bibitem{NGL2011}
   W. G. Newton, M. Gearheart, and B. A. Li, arXiv:1110.4043.

\bibitem{AK1996}
   N. Andersson and K. D. Kokkotas, Phys.\ Rev.\ Lett.\ {\bf 677}, 4134 (1996).

\bibitem{Sotani2001}
   H. Sotani, K. Tominaga, and K. I. Maeda, Phys.\ Rev.\ D {\bf 65}, 024010 (2001).

\bibitem{Sotani2004}
   H. Sotani, K. Kohri, and T. Harada, Phys.\ Rev.\ D {\bf 69}, 084008 (2004).

\bibitem{SYMT2011}
   H. Sotani, N. Yasutake, T. Maruyama, and T. Tatsumi, Phys.\ Rev.\ D {\bf 83}, 024014 (2011).

\bibitem{L1981}
   J. M. Lattimer, Annu.\ Rev.\ Nucl.\ Part.\ Sci.\ {\bf 31}, 337 (1981).

\bibitem{OI2003}
   K. Oyamatsu and K. Iida, Prog.\ Theor.\ Phys.\ {\bf 109}, 631 (2003).

\bibitem{OBG2010}
   F. \"{O}zel, G. Baym, and T. G\"{u}ver, Phys.\ Rev.\ D {\bf 82}, 101301
(2010).

\bibitem{SLB2010}
   A. W. Steiner, J. M. Lattimer, and E. F. Brown, Astrophys.\ J. {\bf 722}, 
33 (2010). 

\bibitem{IS1997}
   K. Iida and K. Sato, Astrophys.\ J. {\bf 477}, 294 (1997). 

\bibitem{RP1994}
   D. G. Ravenhall and C. J. Pethick,  Astrophys.\ J. {\bf 424}, 846
(1994). 

%\bibitem{SHOII1991}
%   T. Strohmayer, H. M. van Horn, S. Ogata, H. Iyetomi, and S. Ichimaru, Astrophys.\ J. {\bf 375}, 679 (1991).

\bibitem{SHOII1991}
   T. Strohmayer {\it et al.}, Astrophys.\ J. {\bf 375}, 679 (1991).

\bibitem{OI1990}
   S. Ogata and S. Ichimaru, Phys. Rev. A  {\bf 42}, 4867 (1990).

%\bibitem{note}
%    The shear modulus on the hadron-quark mixed phase is also suggested \cite{JO2011}.

\bibitem{PP1998}
   C. J. Pethick and A. Y. Potekhin, Phys.\ Lett.\ B {\bf 427}, 7 (1998).

\bibitem{JO2011}
  N. K. Johnson-McDaniel and B. J. Owen, arXiv:1110.4650.

\bibitem{ST1983}
   B. L. Schumaker and K. S. Thorne, Mon.\ Not.\ R. Astron.\ Soc.\ {\bf 203}, 457 (1983).

\bibitem{PCR2010}
   C. J. Pethick, N. Chamel, and S. Reddy, Prog.\ Theor.\ Phys.\ Suppl.\ 
{\bf 186}, 9 (2010).

\bibitem{SNIO}
   H. Sotani, K. Nakazato, K. Iida, and K. Oyamatsu (unpublished).

\bibitem{exp} 
   See, e.g., B. A. Li, C. W. Chen, and C. M. Ko, Phys.\ Rep.\
{\bf 464}, 113 (2008); M. B. Tsang {\it et al.}, Phys.\ Rev.\ Lett.\ 
{\bf 102}, 122701 (2009). 






\end{thebibliography}
\end{document}